\documentclass[12pt,preprint]{aastex}
\received{}
\accepted{}
\slugcomment{}

\begin{document}

\title{The Independence of AGN Black Hole Mass and Radio-Loudness}

\author{Jong-Hak Woo\altaffilmark{1},
C. Megan Urry\altaffilmark{2}}

\begin{abstract}
We revisit the issue of whether radio loudness in AGN is
associated with central black hole mass, as has been suggested
in the literature.
We present new estimates of black hole mass for 295 AGN (mostly radio-quiet), 
calculating their radio loudnesses from published radio and optical fluxes, 
and combine with our previously published values,
for a sample of 452 AGN for which both black hole mass and radio loudness 
(or upper limits thereto) are known. 
Among the radio-quiet AGN, there are now many
black holes with mass larger than $10^{9}$ $M_{\odot}$,
extending to the same high black hole masses as radio-loud AGN
of the same redshifts.
Over the full sample, the black hole masses of
radio-loud and radio-quiet AGN 
span the same large range, $10^6 - 10^{10} M_\odot$.
We conclude that radio loudness in AGN does not depend 
strongly on central black hole mass.
\end{abstract}

\altaffiltext{1}{Department of Astronomy, Yale University, P.O. Box 208101, New Haven, CT 06520-8101; jhwoo@astro.yale.edu}
\altaffiltext{2}{Department of Physics and Center for Astronomy and Astrophysics, Yale University, P.O. Box 208121, New Haven, CT 06520-8121; meg.urry@yale.edu}

\section{Introduction}

The nine out of ten known AGN that are radio-quiet closely resemble the
10\% that are radio-loud.\footnote{We use the standard definition of
radio-loudness, $R>10$, where $R \equiv F_{5~{\rm GHz}} / F_{B}$
(Kellerman et al. 1989).}
The physical origin of this radio-loudness dichotomy remains a mystery.
For this reason, the apparent connection between black hole mass and
radio loudness, reported first by
Franceschini, Vercellone \& Fabian (1998) and apparently confirmed
in subsequent studies (McLure et al. 1999, 
Lacy et al. 2001, Nagar et al. 2002, Jarvis \& McLure 2002)
is potentially very exciting. 
However, other studies found less compelling evidence for a tight connection
(Ho 2002), and revealed a population of radio-loud AGN with quite modest
black hole masses (as low as $M \sim 10^6 M_\odot$; 
Oshlack, Webster \& Whiting 2002).
There was still an apparent threshold effect, wherein
nearly all high-mass black holes ($M > 10^9 M_\odot$) are hosted
by radio-loud AGN (Laor 2000).

In the largest compilation of AGN black hole masses to date, nearly 400 AGN,
including 157 AGN for which radio loudness was measured,
Woo \& Urry (2002; hereafter Paper~I) could not rule out such a threshold effect
because no AGN with $M_{BH} > 10^9 M_\odot$ was radio-quiet. 
In contrast, other trends with black hole mass 
that were previously reported ---
such as $L$ increasing with $M_{BH}$ ---
disappeared in this larger AGN sample; 
apparently sample selection effects in earlier, 
smaller studies created spurious trends.
But the absence of high-mass radio-quiet AGN remained.

Woo and Urry suspected this could be a selection effect,
particularly because the redshifts of AGN with 
$M_{BH} > 10^9 M_\odot$ were considerably higher
than those with lower black hole masses, and 
were from differently selected, not directly comparable
samples. However, the available data were insufficient to 
evaluate this concern.

In the present paper, we ask the simple question,
are there radio-quiet AGN with $M_{BH} > 10^9 M_\odot$?
We look in particular at higher luminosity radio-quiet AGN, 
at higher redshifts than those in Paper~I, 
comparable to the redshifts of the radio-loud AGN with high $M_{BH}$.
In Section 2, we describe the three samples of radio-quiet AGN.
In Section 3 we present the black hole masses and radio loudnesses
for a total of 452 AGN, nearly tripling the previous
sample. Section 4 gives our conclusions.

\section{Black Hole Mass and Radio Loudness for Three AGN Samples}

Most black hole mass estimates in the literature are derived from
optical or UV luminosity, using the correlation between size of
the broad-emission-line region and luminosity (Kaspi et al. 2000,
McLure \& Jarvis 2002, Vestergaard 2002). We use this method, 
combining luminosity with broad emission line width. 
Therefore, the search for high-mass black holes in AGN 
corresponds to a search for high-luminosity AGN 
with measured line widths. 
Three large quasar samples,
described below, are useful for our purpose.
This collection of different samples is not
useful for inferring the intrinsic distribution of $M_{BH}$, 
since the selection criteria vary from one sample to the next,
but it is sufficient to determine whether high-$M_{BH}$ exist in
radio-quiet AGN, at redshifts comparable to those of the
high-$M_{BH}$ in radio-loud AGN.

\subsection{The BQS Sample} 

The Bright Quasar Survey (BQS) sample is
a subset of the PG quasar sample (Schimidt \& Green 1983).
Boroson \& Green (1992) present detailed emission-line properties 
for the 87 low-redshift sources ($z < 0.5$) of the 114 BQS AGN,
and other characteristics of the low-redshift BQS, including
continum energy distributions (Neugebauer et al. 1987) and 
radio properties (Kellerman et al. 1989), are also
in the literature.

For the 87 AGN in the Boroson \& Green (BG) sample,
we collected or made black hole mass estimates.
Seventeen are known from reverberation mapping (Kaspi et al. 2000),
at present the most accurate method for estimating black hole mass,
so these we simply adopted. 
For the remaining 70 AGN, we use 
optical luminosity and $H_{\beta}$ line width
assuming the broad line region is virialized by the central black hole 
(Eqn.~3 of Paper~I), 
correcting the published luminosity as needed according to our 
adopted cosmology ($H_0 = 75$~{\rm km/s/Mpc}, $q_0 = 0.5$, $\Lambda = 0$). 
We note that 
7 of these were previously estimated by McLure \& Dunlop (2001)
but with a different cosmology;
Laor (2000) plotted all 87 in his Fig.~2 but did not tabulate the masses; 
and 24 were previously published in our Paper~I.

The estimated black hole masses of the 70 radio-quiet AGN range from roughly $10^{6}$
to $10^{9}$ $M_{\odot}$, with a few larger than $10^{9}$ $M_{\odot}$.
Figure~1 plots radio loudness for this sample,
also taken from Boroson \& Green (1992), 
versus black hole mass.
If the black hole mass function is steep,
the volume represented by the low redshift range of this sample
($z<0.5$) may be too small to contain many large black holes;
nonetheless, it is clear there do exist radio-quiet AGN with 
$M_{BH} > 10^{9}$~$M_{\odot}$.

\subsection{McIntosh Sample}

To find the possibly rare high-mass black holes, we looked for
a higher-redshift sample of radio-quiet AGN for which sufficient
data exist to estimate black hole masses.
The H-band spectroscopic study by McIntosh et al. (1999) 
of 31 high-redshift AGN ($2 < z < 2.5$),
17 radio-quiet and 14 radio-loud,
includes rest-frame $H_{\beta}$ line widths and 
rest-frame V-band luminosities.
We used the same method 
(optical luminosity and $H_{\beta}$ line FWHM) to estimate black hole mass, 
after correcting the luminosities to our cosmology. 
Black hole masses range from a few to ten billion $M_{\odot}$. 
All radio-quiet AGN in this sample have $M_{BH} > 10^9 M_\odot$.

\subsection{LBQS Quasars}

Our previous study (Paper~I) found
no radio-quiet AGN with black hole mass larger than $10^{9}$ $M_{\odot}$;
however, most of the larger black hole masses in that paper were for 
(radio-loud) AGN in the redshift range $0.5<z<1.0$, 
compared to a lower redshift range, $z<0.5$, for the radio-quiet AGN.
To eliminate this possible selection effect,
we here look specifically for radio-quiet AGN at $0.5 < z < 1.0$.

The Large Bright Quasars Survey (LBQS) contains 
1058 optically-selected AGN at redshifts $0.2<z<3.3$
(Hewett, Foltz \& Chaffee 1995). 
Emission line properties for the LBQS are published in Forster et al. (2001).
Since the $H_{\beta}$ line is not available for the most AGN in the 
redshift range of interest, we instead use the MgII line.
McLure and Jarvis (2002) showed there is a correlation between
the reverberation-mapped size of the Mg~II-emitting region and
the UV luminosity (at 3000 \AA); we re-derived this relation for
our cosmology. 
Assuming isotropic orbits, we have:
 
\begin{equation}
M_{BH} = 4.2~ (\frac{\lambda L_{\lambda}(3000{\rm \AA})}{10^{44}~{\rm ergs/s}})^{0.47} ~({\rm FWHM})^{2} ~.
\label{Luv}
\end{equation}

\noindent
UV luminosities for the 251 LBQS AGN at $0.5 < z < 1.0$
were calculated from the spectral indices and measured fluxes reported by Forster et al. (2001).

To obtain radio luminosities, we searched the Faint Imaging Radio Source (FIRST) survey, 
which covers roughly a half of the LBQS area (Becker, White, \& Helfand 1995). 
Two-hundred one out of the 251 AGN 
at $0.5<z<1.0$ are covered by FIRST survey; 
16 are FIRST radio sources (Hewett, Foltz \& Chaffee 2001).
Since the other 185 AGN are not detected, we derived radio upper limits at 5~GHz
by converting the 1~mJy flux limit at 1.4~GHz using a spectral index $\alpha=0$. 
Optical luminosities were calculated from the spectral indices 
and measured fluxes reported by Forster et al. (2001).

\section{Black Hole Mass with Radio Loudness}

In Figure 1, we present new black hole masses and radio loudnesses (or 
upper limits) for 295 AGN, and combine them with the 
157 AGN discussed previously in Paper~I. 
Just as there are radio-loud AGN with small black hole masses, 
there are clearly radio-quiet AGN with large black hole masses.
For $M_{BH}  > 10^9 M_\odot$, there is no statistically significant
difference between the distributions of black hole mass for radio-loud 
and radio-quiet AGN (with 96\% probability that the distributions are
drawn from the same parent distribution, according to a KS test).

This is actually surprising since, 
given the heterogeneity of the overall sample, there is no
reason to expect the {\it distributions} of black hole mass in
radio-loud and radio-quiet AGN to agree.
It indicates more similarity than we have a right to expect. 
The radio-loud AGN were selected primarily at radio-wavelengths, 
while the radio-quiet AGN come from optical surveys, so 
the flux limits and the volume sampled are quite different.
(Sizeable collections of
optically-selected radio-loud AGN will be possible with large surveys
such as the Sloan Survey, but have not previously been available 
because of limited sample size.) 

If black hole mass is independent of radio loudness,
then however heterogeneous the sample selection,
it must be the case (as long as redshift ranges are similar) that
the black hole masses of radio-loud and radio-quiet AGN 
span the same range.
Indeed, for radio-quiet and radio-loud AGN with $M_{BH} > 10^9 M_\odot$,
the redshift ranges are quite similar (although the actual distributions 
do differ at the 99\% confidence level according to a KS test, reflecting
the different selection biases).
Thus the volume and epoch sampled for the new radio-quiet AGN 
are similar to that of the high-$M_{BH}$ radio-loud AGN discussed in Paper~I. 
Once this condition of compatibility is satisfied, 
the difference in black hole mass --- 
i.e., the apparent lack of high-$M_{BH}$ among the radio-quiet AGN 
--- goes away, and
the distributions of black hole mass above $10^9 M_\odot$
are consistent with being the same, independent of radio loudness.

\section{Conclusions}

We conclude that there is no evidence that radio loudness
depends on black hole mass. There does not appear to be
any correlation between $R$ and $M_{BH}$, nor is there
a threshold such that large black hole masses, 
$M_{BH} > 10^9 M_\odot$, are preferentially radio-loud.
The origin of radio loudness remains a mystery.

\acknowledgements

Support for proposal \#HST-GO-09223.01-A 
was provided by NASA 
through a grant from the Space Telescope Science 
Institute, which is operated by the Association of
Universities for Research in Astronomy, Inc., 
under NASA contract NAS 5-26555.

\clearpage
\begin{figure}
\plotone{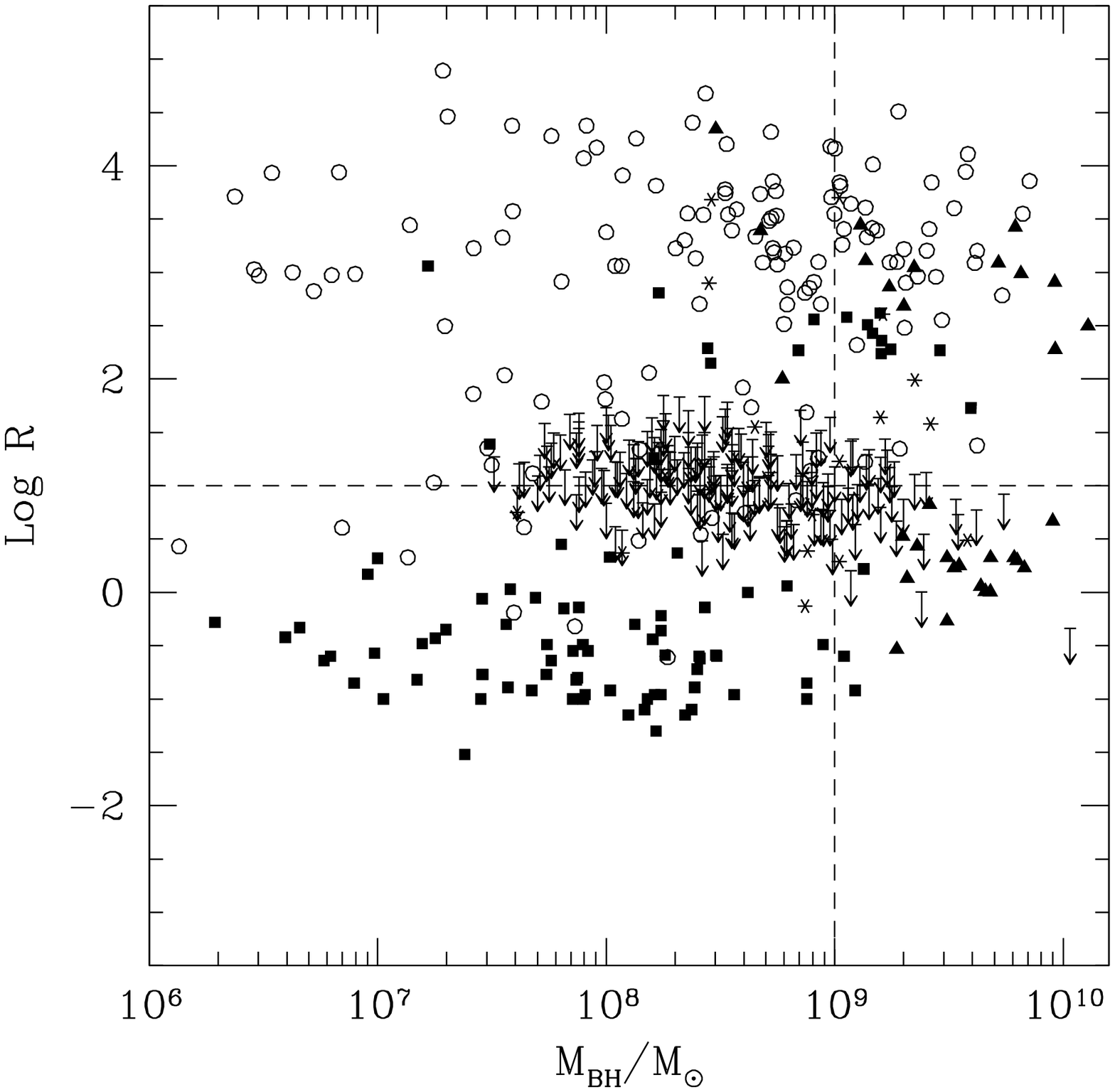}
\caption{Radio loudness versus black hole mass for 452 AGN,
most at $z \lesssim 1$, including
295 new estimates for the samples discussed in \S~2, 
and 157 previously published in Paper I.
There is no correlation between radio loudness and black hole mass.
In particular, at high mass ($M > 10^9 M\odot$) there
are similar distributions of black hole mass
for radio-loud ($R>10$) and radio-quiet AGN.
{\it Squares:} PG quasars, all at $z<0.5$;
{\it circles:} remaining AGN from Paper I, at $0<z<1$;
{\it triangles:} high-redshift quasars ($2<z<2.5$) from McIntosh et al. (1999);
{\it stars:} LBQS quasars, $0.5<z<1$;
{\it arrows:} upper limits for LBQS quasars, $0.5<z<1$.
Radio fluxes for AGN from the Parkes sample of Oshlack et al. (2002)
have been revised downward to account for relativistic beaming, following
Jarvis \& McLure (2002); this was not done in Paper~I.
}
\end{figure}

\end{document}